\documentclass[fleqn,usenatbib]{mnras}

\usepackage[T1]{fontenc}
\usepackage{ae,aecompl}

\usepackage{graphicx}	
\usepackage{amsmath}	
\usepackage{amssymb}	

\title[The 2VT and the ADR]{The two-component virial theorem and the acceleration-discrepancy relation}

\author[C. C. Dantas et al.]{
Christine C. Dantas,$^{1}$\thanks{E-mail: christineccd@iae.cta.br}
Andr\'e L. B. Ribeiro,$^{2}$\thanks{E-mail: albr@uesc.br}
and Hugo V. Capelato$^{3,4}$\thanks{E-mail: hcapelato@gmail.com}
\\
$^{1}$Instituto de Aeron\'autica e Espa\c co (IAE),  
Departamento de Ci\^encia e Tecnologia Aeroespacial (DCTA), \\
S\~ao Jos\'e dos Campos, 12228-904, SP, Brazil\\
$^{2}$Departamento de Ci\^encias Exatas e
Tecnol\'ogicas, 
Universidade Estadual de Santa Cruz, 
Ilh\'eus, 45650-000, BA, Brazil\\
$^{3}$N\'ucleo de Astrof\'{\i}sica Te\'orica, 
Universidade Cruzeiro do Sul, 
S\~ao Paulo, 01506-200, SP, Brazil \\
$^{4}$Divis\~ao de Astrof\'{\i}sica (INPE-MCT), 
S\~ao Jos\'e dos Campos, 12227-010, SP, Brazil
}

\date{Accepted XXX. Received YYY; in original form \today}

\pubyear{2017}

\begin{document}
\label{firstpage}
\pagerange{\pageref{firstpage}--\pageref{lastpage}}
\maketitle

\begin{abstract}
We revisit the ``two-component virial theorem'' (2VT) in the light of recent theoretical and observational results related to the ``dark matter''(DM) problem.  This modification of the virial theorem offers a physically meaningful framework to investigate possible dynamical couplings between the baryonic and DM components of extragalactic systems. In particular, we examine the predictions of the 2VT  with respect to the ``acceleration-discrepancy relation'' (ADR). Considering the combined data (composed of systems supported by rotation and by velocity dispersion), we find that: (i) the overall behavior of the 2VT is consistent with the ADR; and (ii) the 2VT predicts a nearly constant behavior in the lower acceleration regime, as suggested in recent data on dwarf spheroidals. We also briefly comment on possible differentiations between the 2VT and some modified gravity theories.
\end{abstract}



\section{Introduction \label{INTRO}}

A fundamental question currently spanning astrophysics, cosmology and particle physics is the ``dark matter problem''. It arose in the 30's as a curious disagreement in the mass estimates of astronomical bodies, and persists today as one of the greatest unsolved problems in physics \citep{Ber16, Gas16,Fre17}. Dynamical mass estimates of galactic systems exceed those obtained from their luminous contributions. Such a discrepancy would in principle be solved by conjecturing the existence of some additional mass (undetectable in terms of electromagnetic emission): its contribution would then account for the gravitational binding of galaxies and clusters of galaxies.

This initial puzzle deepened in the past decades, with the improvement of observational techniques, bringing forth higher quality surveys and the gravitational lensing methods \citep{Tre10}, expanding our studies of extragalactic systems and large scale structures. At the same time, the accuracy of cosmological parameters improved \citep{Pla16a},  requiring not only a  ``non-baryonic dark matter'' (DM), but also a dominant ``dark energy'' (DE) component, in order to account for an apparent accelerated cosmic expansion of the Universe. These investigations established a ``standard model'' of Big Bang cosmology: the ``Lambda Cold Dark Matter''  ($\Lambda$CDM) model, providing a relatively consistent picture to describe various independent properties of the Universe (for implications of recent Planck data regarding a few tensions with independent results, see \citealt{Pla16a}; and with respect to alternative scenarios, see  \citealt{Pla16b}).

The Virial Theorem (VT; c.f. \citealt{Bin87}) is one of the main methods to estimate the total mass of a galactic system. In Zwicky's pioneering paper  \citep{Zwi37}, the VT was used to estimate the total mass of the Coma cluster, under the assumption that this system is in stationary equilibrium, at least in a first approximation (as inferred from its regular, spherically symmetric distribution).  This work not only marks the beginning of the ``missing mass problem'', but also highlights the importance of the VT as a fundamental tool in extragalactic astrophysics, particularly in the context of this problem  (for an historical account on virial mass estimates of galaxy clusters, see, e.g., \citealt{Biv00}). The application of the VT is not straightforward, as it requires an underlying hypothesis of stability of the system and well-understood observational selection effects (a pioneer study in this regard is given in \citealt{Aar72}).  More recently, gravitational lensing mass estimates have shown compatibility with the idea that the overestimation of total virial masses can be attributed by the presence a dominant DM component \citep{Tre10}. 

The VT refers to a global condition on the kinetic and potential energies of the system as a whole. But it is also a matter of great interest to gain further information on the detailed equilibrium requirements of a cluster composed of other subsystems, such as a less massive luminous component embedded in a dominant halo. In 1959, \citet{Lim59} first derived a more general, two-component VT form, in order to model a cluster embedded in an extended (nonviscous) gaseous background, interacting only through gravity. Clearly, Limber's  ``two-component virial theorem'' (2VT) can also be applied to dark matter halos, by re-interpreting the ``extended gaseous background'' as a dark matter halo. In this regard, \citet{Smi80} designated the  ``Limber effect'' the overestimation of the total mass, obtained from the application of the (usual) VT, for systems with an extended, unseen background. 

In certain VT applications, constraints on the dark matter halo can be obtained. For example, when the stellar contribution to the gravitational field can be considered sufficiently small in comparison to the dark matter component, so that the former is primarily moving in the gravity field of the dark matter halo. Thus, for such a tracer stellar populations, the tensor VT \citep{Bin87} could be used to constrain the dark matter halo in our Galaxy \citep{Agn12a}  and in the dwarf spheroidal galaxy Sculptor \citep{Agn12b}. By extending the tensor VT to subsystems, more information can be obtained about individual components, than that acquired by the application of the usual VT to the system as a whole  \citep{Bro83, Cai84, Cai92}. For instance, the structural configuration of a component in equilibrium may be distorted by the tidal force induced by the other, introducing a length dependence on the baryonic subsystem induced by the dark matter halo \citep{Mar03}. These, and possibly other dynamical effects, could provide, at least partially, a regulatory mechanism for explaining tight observational constraints, such as the ``Fundamental Plane'' of elliptical galaxies \citep{Djo87, Dre87,Cap95, Dan00, Dan03,DOn16}, including a general, combined observational effect comprising a large range in scales and different types of systems, the so-called ``Cosmic Metaplane'' \citep{Bur97,Dan00,Sec00}. 

A recent observational result of particular interest  is the existence of a tight correlation between the radial acceleration derived from rotation curves of galaxies and the observed distribution of baryons (hereon, the ``Acceleration Discrepancy  Relation'', ADR, \citealt{McC16,Lel17}). This empirical relation suggests a strong coupling between dark and baryonic components, possibly related to galaxy formation mechanisms. The 2VT provides a physically meaningful framework to investigate dynamical couplings between these components, based on their mutual equilibrium conditions. Note that the usual (one-component) VT does not address any systematic couplings between ``hidden'' and  baryonic masses, it just implies that a ``remainder'' mass must be added, by contingency, to the dynamical equilibrium budget of the system. But the 2VT formulation indicates a correction that depends systematically on the dark component in which the baryonic mass is embedded. 

In this paper, we revisit the 2VT to address these recent theoretical and observational results. Our paper is outlined as follows. In Sec. 2 we present the 2VT, in terms of a suitable expression to fit the data in the ADR space.  In Sec. 3, we compare the 2VT predictions with data for Late Type Galaxies (LTGs) and Dwarf Spheroidals (dSphs) and discuss the ``flattening'' behaviour of dSphs, as indicated in recent data. Finally, we briefly comment on possible differentiations between the 2VT and
some modified gravity theories. In Sec. 4 we present our conclusions. 

The usual, one-component, VT will be denoted here as ``1VT'' (c.f. Eq. \ref{Eq:1VT} in App. \ref{Ap:Re2VT}). Our notation uses the index $B$ to refer to the baryonic matter and $D$, to the dark matter component; ${\rho}$, the respective average matter density within a given radius $r$.


\section{The two-component virial theorem \label{Sec:models}}

In this section, we present the ``two-component virial theorem'' (2VT, \citealt{Lim59,Dan00,Sec00}) in a suitable form to be compared with the ``Acceleration Discrepancy  Relation'' (ADR, \citealt{McC16,Lel17}). In App. \ref{Ap:Re2VT}, we review the 2VT as originally derived  in \citet{Dan00} and present details on its derivation.

The 2VT provides a correction term to the 1VT,  which accounts for the influence of a second component (the putative DM halo). In terms of acceleration variables, the observed acceleration, as predicted by the 2VT, is given by (c.f. App. \ref{Ap:Re2VT}):
\begin{equation}
g_{\rm obs, 2VT} = g_B + Rg_D .  \label{Eq:2VTsum}
\end{equation}
\noindent where $g_B$ is the radial acceleration of the baryonic component, $g_D$ is the radial acceleration associated with the dark component (c.f. Eq. \ref{Eq:gB}), and $R$ is a parameter depending only on the properties of the baryonic matter distribution, relating the projected (2D) to the  ``physical'' (3D) radii of the observed baryonic component. The 1VT is, simply:
\begin{equation}
g_{\rm 1VT} = g_B.  \label{Eq:1VTg}
\end{equation}

The 2VT gives a simple linear correction to the 1VT, and the behaviour of the 2VT as seen in a $\log$--$\log$ plot appears as curved line. In App. \ref{Ap:2VTfit}, we present a brief illustration of the 2VT as seen in that plot, showing how variations in the parameters $(g_D, R)$ affect it.  Here we  highlight two relevant facts:

\begin{itemize}
\item{The 2VT curve with a fixed $(g_D, R)$ represents a parametrized family of baryonic--DM systems within an arbitrary baryonic mass range.}
\item{The 2VT predicts some discriminating characteristics in the $\log$--$\log$ plane: {\it (i)} a bending departing from the 1VT and {\it (ii)} a nearly constant behavior in the lower acceleration regime. The parameters $(g_D, R)$ cannot modify these features, only allowing for an adjustment of the height of the asymptotic behavior in the lower acceleration regime, or, alternatively, the point at which the 1VT is retrieved.}
\end{itemize}

In other words, even with two free parameters, the 2VT presents a relatively ``rigid'' form, implying that this model cannot be arbitrarily contrieved to fit the data. This level of predictive power is an important feature of the 2VT.


\section{Observational comparisons \label{RE-2VT}}

In this section, we analyse the 2VT in terms of current observational data leading to the so-called ``acceleration-discrepancy relation'' (ADR; \citealt{McC16, Lel17}).

\subsection{The acceleration-discrepancy empirical relation} \label{sub:ADR}

The ADR is an empirical best-fit relation for the centripetal acceleration estimated from the rotation curves of rotationally-supported galaxies, and it was first described by  \citep{McC16}: 

\begin{equation}
\mathcal{F}(g_B) = {g_B \over 1 - e^{- \sqrt{g_B/a_M}}}. \label{Eq:empirical}
\end{equation}
\noindent Note that the above relation tends to a linear slope at high accelerations and $g_{obs} \propto \sqrt{g_B}$ at low accelerations. 

The inclusion of dwarf spheroidals (dSphs) data \citep{Lel17}, however, seem to imply a flattening of the above relation, specially for the ultrafaint dSphs. The present data is not conclusive, but this flattening behavior seems to be favored at this time, and suggested a modification of the relation above  to \citep{Lel17}:

\begin{equation}
\mathcal{F}_m(g_B) = {g_B \over 1 - e^{- \sqrt{g_B/a_M}}} + \hat{g}e^{- \sqrt{g_Ba_M/\hat{g}^2}}, \label{Eq:empirical2}
\end{equation}
\noindent as a better fit to the data, for the tendency of dSphs to deviate from the original ADR (Eq. \ref {Eq:empirical}).

\subsection{The 2VT and the ADR} \label{sub:2VT-ADR}

\begin{figure} 
\includegraphics[width=1\columnwidth]{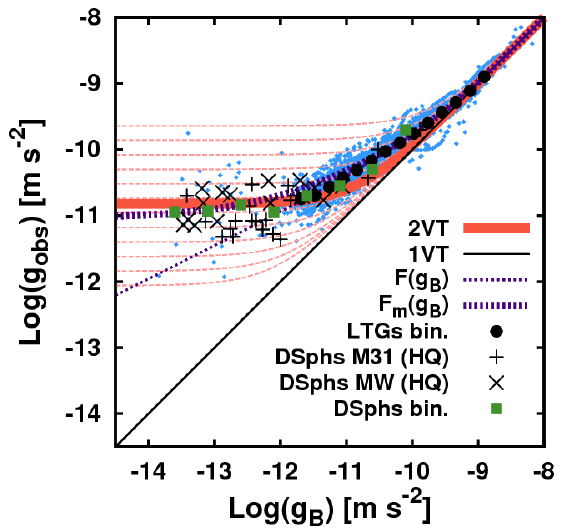}
\caption{ (Color online). The acceleration obtained from the baryonic matter ($g_{\rm B}$) vs. the observed acceleration ($g_{\rm obs}$). The 2VT predicts that $g_{\rm obs}$ obeys Eq. \ref{Eq:2VTsum}. We show a series of 2VT curves (thin dashed lines) covering most of the data, using different $R$ values, in equal (dex) steps, from $R = 1.905$ (upper) to $R = 0.007$ (lower curve), with $g_D$ fixed to the Milgrom $a_M$ scale (see main text). An  ``eyeball'' best 2VT curve for the data as a whole is also indicated (thick continuous line), with $R=0.125$. Empirical relations  expressed by Eq. (\ref{Eq:empirical}) (labeled ``$F(g_B)$''; thin dotted line) and Eq. (\ref{Eq:empirical2}) (labeled ``$F_m(g_B)$''; thick dotted line), with the respective best fit parameters obtained in \citet{Lel17}, are shown for comparison. Data is taken from \citet{Lel17} and corresponding binned data is shown in larger symbols, as indicated in the legend. ``High quality'' (HQ) data for dSphs (M31 and Milky Way) is highlighted with different symbols. Small dots in the background represent the full dataset for both LTGs and dSphs. The line of unit ($g_{\rm obs} = g_B$) expresses the 1VT expectation for baryons only.\label{Fig:compare}}
\end{figure}

In Fig. \ref{Fig:compare}, we reproduce the suggested ADR best-fit forms, Eqs. \ref{Eq:empirical} and \ref{Eq:empirical2}, together with  a series of 2VT curves, Eq. \ref{Eq:2VTsum}, for illustrative purposes. We provisionally fix $g_D$ to Milgrom's acceleration scale \citep{Mila},
\begin{equation}
a_M \equiv 1.2 \times 10^{-10} ~[m~s^{-2}], \label{milgrom_am}
\end{equation}
\noindent and use different values of $R$, covering most of the data (represented by small dots in the figure; binned data is indicated by larger symbols). Detailed fits are presented below (see also App. \ref{Ap:2VTfit}).

The data was taken from \citep{Lel17}, which includes rotationally suported systems (late-type galaxies, LTGs, from SPARC data) and dwarf spheroidals (dSphs), which are suported by velocity dispersion. We indicate the subset of the ``high quality'' (HQ) data which, among other cases regarding both the quality of the velocity dispersion determination as the sphericity of the dSphs, excludes all those strongly affected by the tidal forces produced by their host galaxies (see \citealt{Lel17} for details).
It is interesting to compare our figure with Fig. 3 of  \citet{McC16} and figures 10, 11 and 12 of \citet{Lel17}.  

LTGs are preferentiably located nearby the transition scale $a_M$ between the Newtonian and MONDian (''MOdified Newtonian Dynamics''; \citealt{Mila,Milb,Milc}) regimes, whereas dSphs are mainly located at lower acceleration regimes (see also Sec. \ref{sub:modgrav} below). The ultrafaint dSphs particularly contributes to the data in the lower acceleration region. As Fig. \ref{Fig:compare} shows, a band of 2VT curves cover in the right sense the ADR of rotationally-supported galaxies, including the low-acceleration flattening region, indicated by ultrafaint dwarf spheroidals.

\begin{figure} 
\includegraphics[width=1\columnwidth]{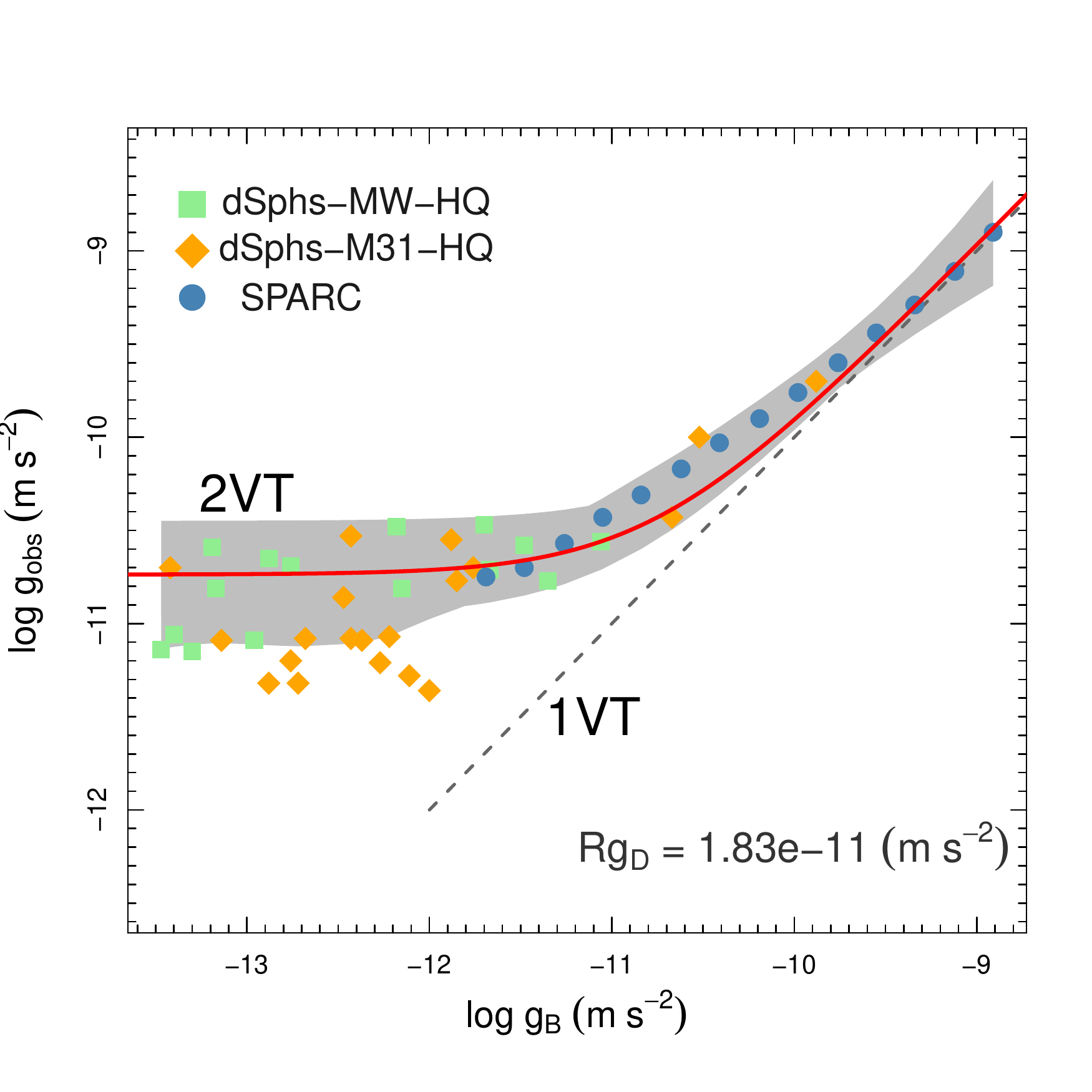}
\caption{(Color online). Same as Fig. \ref{Fig:compare}, using binned LTG data and HQ dSphs data, with a 2VT best-fit curve obtained from a linear regression method, including the $95$\% confidence band around the regression fit (grey band), as explained in the main text. Fixing  $g_D$ to Milgrom's acceleration scale (Eq. \ref{milgrom_am}) gives $R=0.152$.\label{Fig:detail-fit1}}
\end{figure}

We also provide a quantitative fit to the 2VT (Fig. \ref{Fig:detail-fit1}), using the binned data for LTGs and the ``high quality'' (HQ) subset dSphs data, obtained with linear regression, a procedure that allows us to find the value for the parameter $\mathcal{G} \equiv Rg_D$ (c.f. Eq \ref{Eq:2VTsum}), with respective standard errors, minimizing the sum of the squares of the data points distances to the curve given by the 2VT. We ran the code $lm$ under the stats package in R \citep{RMan14}. The code should be able to determine the best-fit parameter regardless of the initial guess. However, to avoid convergence problems, we created a broad grid that encloses reasonable values for the parameter $\mathcal{G}$ (following approximately the ranges obtained in  Fig. \ref{Fig:compare}). We also computed  the residual standard deviation at each point on the grid to find the best parameters choice. The best-fit value found was $\mathcal{G} = 1.83 \times 10^{-11}$ (p-value: $p = 3 \times 10^{-4}$); fixing  $g_D$ to Milgrom's acceleration scale (Eq. \ref{milgrom_am}) gives $R=0.152$.

\begin{figure}
\includegraphics[width=1\columnwidth]{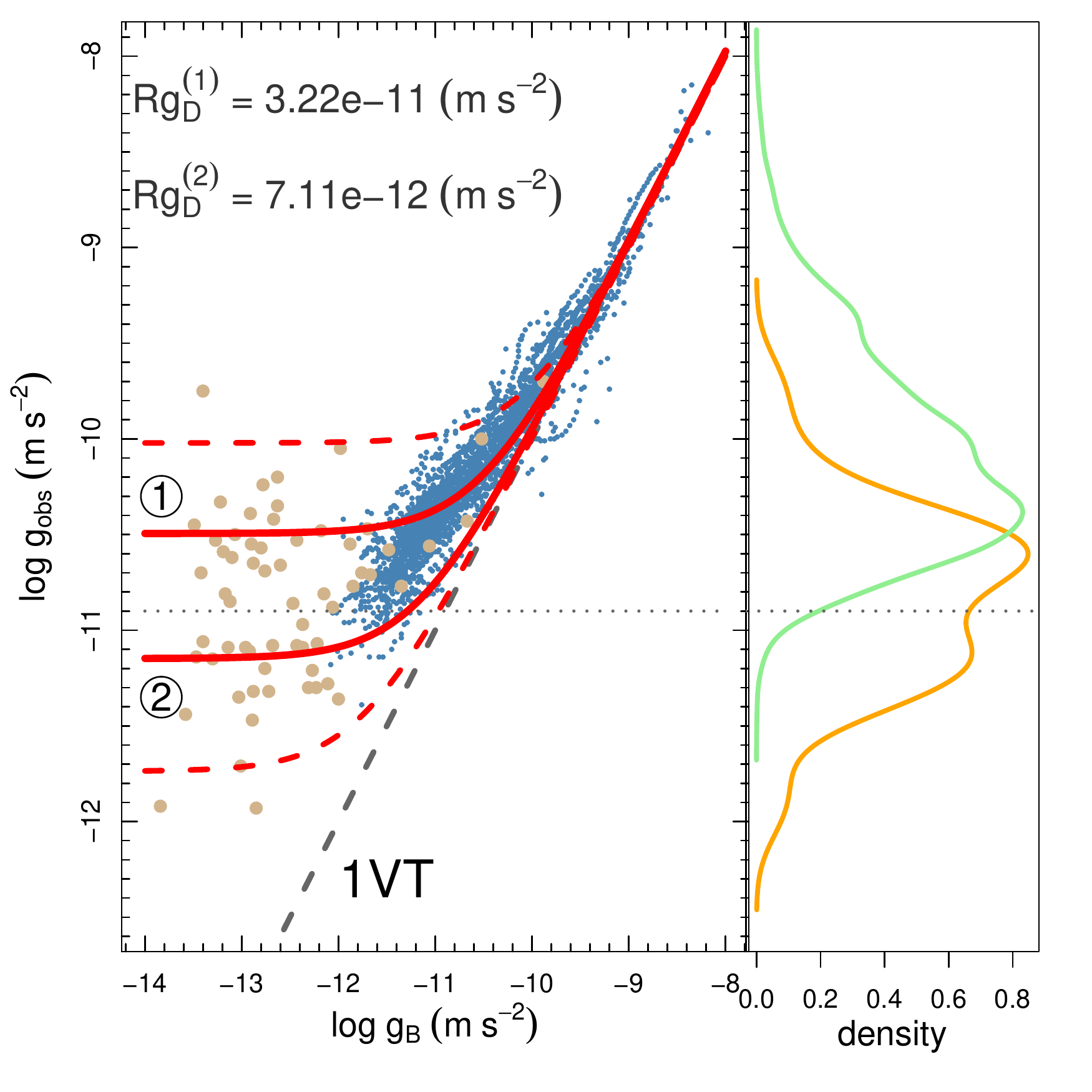}
\caption{(Color online). {\it Left panel:} Same as Fig. \ref{Fig:detail-fit1}, using the full (unbinned) LTG (small dots) and dSphs (circles) data, with detailed 2VT best-fits using a linear regression method, including the 95\% confidence band around the regression fit (dashed curves), as explained in the main text. {\it Right panel:} density distributions in the data, shown separately for LTGs and dSphs, with a possible bimodality detected in the dSphs data (indicated by an horizontal dashed line).  Best-fit values are shown in the legend.\label{Fig:detail-fit}}
\end{figure}

We also present a brief analysis based on the full data for (unbinned) LTGs and dSphs (Fig. \ref{Fig:detail-fit}), based on the corresponding density distributions in their data. A possible bimodality in the dSphs data is indicated   (the separation of modes is represented by a horizontal line in the figure). A best-fit curve was obtained separately for each individual mode (above and below this line), giving two different sets for the best-fit parameter $\mathcal{G} \equiv Rg_D$, as indicated in the legend of Fig. \ref{Fig:detail-fit}. The $95$\% confidence upper limits are the upper envelope for the regression fit obtained in the upper mode, and  the lower envelope in the lower mode. The p-value of the F test for mode 1 is $p < 10^{-4}$, whereas the fitting for mode 2 does not converge easily, giving a high p-value, $p = 0.1873$. In other words, the 2VT fits mode 1 very well, but not mode 2. In this latter case, the 2VT curve in the high acceleration range misses the large volume of LTGs data in order to contemplate the lower acceleration data for dSphs. 

It is interesting to note that mode 2 is mainly composed of M31 dwarf spheroidal data.
Indeed, Walker et al. (\citealt{Wal10} and references therein) finds a systematic difference between M31 and MW dSphs velocity dispersions at a given half-light radius, with the former having lower velocity dispersions than the later.  It is not clear whether this effect is the result of some systematic bias or an intrinsic signature of different formation processes in these sytems. A more detailed analysis of this possible bimodality in the context of the 2VT is left for a future work.

\subsection{The ``flattening'' behavior of Sphs}

In the section 8.3.2 in \citet{Lel17} (``New physics in the dark sector?''), some considerations were given in order to account for the ``flattening'' ADR behavior in dSphs. In particular,  the internal gravitational field of some of these systems could be ``contamined'' with that of their  host galaxy, even though the ``quality cut'' mentioned previously was adopted in the data. An attempt to bring the dSphs to the extrapolated behavior of the LTGs (their Fig. 14) is not in accordance with the standard MOND modifications for external field effects and seem to be not well understood in the MOND framework \citep{Lel17}  (see also next section). 

The 2VT offers a natural explanation for this ``flattening'', since the correction term to the observed acceleration does not depend on $g_B$, but on the product $Rg_D$  (Eq. \ref{Eq:2VTsum}), which is fixed for a given family of baryonic--DM systems. That is, the 2VT cannot be much contrieved, as illustrated in the App. \ref{Ap:2VTfit}. Therefore the somewhat ``mysterious'' dwarf spheroidal ``flattening'' behavior, if observationally confirmed, is a prediction of the 2VT, and such a prediction cannot be fine-tuned.

One possibility is that this flatening is due to tidal effects from the host \citep{Cai92}. In this case,  the virial theorem should be formulated with the presence of a ``third component'' (host) in supposed equilibrium with the coupled baryonic-DM system. However, the resulting corrections would increase in complexity and become cumbersome, except if assuming some special cases and/or symmetries \citep{Cai92}. Evidently, it would have to be assumed that the dSph system is indeed in virial equilibrium with the host. Given the good agreement of the 2VT with the data, we consider that any contribution of a host within the gravitational radius of the baryonic system is negligible, or otherwise may contribute only to  disperse of the data around the flatenning region (low acceleration limit).  Disentangling such possible contributions from others, such as dissipative effects \citep{Rib10}, are difficult to address at this point. LTGs, being located at higher acceleration regimes, are not affected by such considerations.

\subsection{Systems with variable $(g_D,R)$}

The structural parameter $R$ depends on the type of galaxy, given their different structural equilibrium configurations, reflected on their different shapes. The value of $g_D$ may also vary depending on the DM halo contribution within the baryonic gravitational radius, which may have different scalings for different systems. In the 2VT prediction (Eq. \ref{Eq:2VTsum}), these variations should be independent, but in a more general formulation (see App. \ref{Ap:Limber}, \citealt{Lim59}), the equivalent to our $R$ parameter would depend on the DM halo distribution as well. These variations would produce a spread around a given unique 2VT curve (i.e., around a family baryonic--DM systems), which would tend to be more obvious in the very low acceleration regime, where the 2VT curve admits a dominance of the DM halo (evidently, depending on the combined product, $Rg_D$).

\subsection{A note on the 2VT and modified gravity theories } \label{sub:modgrav}

A broad alternative scenario to the DM paradigm considers that at large scales gravity should be modified \citep{Cal17}. A proposal in this direction, initially applied to galaxies and clusters of galaxies, was provided by Milgrom in the 80's as a modification of Newton's law in the extremely weak field regime (\citealt{Mila,Milb, Milc}; c.f. reviews in \citealt{Fam12,Mil15}). There are currently several implementations of MOND, but the basic mechanism is a departure from Newtonian dynamics below the critical acceleration, $a_M$, Eq. \ref{milgrom_am}. A connection of MOND with cosmology was also considered by \citet{Mil99}, where this acceleration scale could arise from a vacuum effect. 

Another testable proposal has been recently given by Verlinde, who states that gravity itself might arise as an entropic force, so that spacetime emerges from a microscopic substratum \citep{Ver11,Ver16}. Verlinde's proposal has a close connection with previous works, aiming to derive general relativity from thermodynamics \citep{Jac95, Pad10}. A question was then brought forth concerning a possible retrieval of MOND-like behavior from entropic arguments (see \citealt{Mil16} and references therein), so that these ideas have been taken, separately or combined, as possible alternative candidates to the DM paradigm.

MOND has been subjected to a large number of observational tests throughout many years (see reviews above), and a few tests for Verlinde's theory have been made (e.g., \citealt{Par17}, \citealt{Hos17} and references therein), but results seem premature at this point, specially due to simplifications adopted in the theory. A covariant proposal for Verlinde's theory, along with some clarifications, has recently been made, which may advance the theory and favor cleaner observational testing \citep{Hos17}. 

The ADR is a particularly interesting test for these alternative theories, specially on the possibility of differentiating DM from non-DM based models. Qualitatively, it is possible that MOND reproduces the ADR (see, e.g., discussion in \citealt{Lel17}), although it is unclear at this point whether the behaviour of dSphs at the lower acceleration region can be accounted for. Therefore, at this point, a detailed prediction of MOND for that specific lower acceleration region ($g_B < -12$ dex) is necessary for a clear comparison with the 2VT prediction; however, to the knowledge of the authors, such a prediction is not yet available. 

As for Verlinde's theory, the main phenomenological result has been derived from spherical symmetry (Eq. 7.47 in \citealt{Ver16}):
\begin{equation}
{\rho}^2_D (r)= \left [4 - \bar{\beta}_B(r) \right ] {a_0 \over 8 \pi G} {{\rho}_B(r) \over r}, \label{Eq:Ver}
\end{equation}
\noindent where $a_0 = 6 a_M$, and $\bar{\beta}_B(r)$ is the slope parameter, $\bar{\beta}_B(r) = - {d \log {\rho}_B(r) / d \log r}$, and subscript $D$ refers to the resulting ``apparent'' dark matter effect.
\noindent By the use of Eqs. (\ref{Eq:gB}), (\ref{Eq:rhoave}) of App. \ref{Ap:Re2VT} into Eq. (\ref{Eq:Ver}), the latter can be written in terms of acceleration variables as:
\begin{equation}
g_{\rm (D, VEGT)} =  \left [4 - \bar{\beta}_B(r) \right ]^{1\over 2}
 \sqrt{a_Mg_B}.   \label{Eq:Ver-g}
\end{equation}
\noindent Hence, on general grounds and using simplifying assumptions, Verlinde's theory predicts a  correction $g_{\rm (D, VEGT)}$ for the observed acceleration $g_{\rm obs}$  that depends on $\sqrt{a_Mg_B}$, similarly to MOND (e.g. \citealt{Lel17}), with prefactors that may differ. On the other hand, the 2VT predicts that this correction depends on the product $Rg_D$ (c.f. Eq. \ref{Eq:2VTsum}), that is, it is not a function of $g_B$, as in the former theories, but a function of the baryonic matter distribution ($R$) and the acceleration scale ($g_D$ within $r_B$) associated with the DM component of the system.

This difference in the functional dependence of the correction term for the observed acceleration imposes a qualitative distinction on the form of the predicted ADR. In Verlinde's theory and in MOND, obtaining a nearly constant behaviour in the lower acceleration regime  requires a functional tuning (in terms of $g_B$) in their respective correction terms. On the other hand, the 2VT already predicts an asymptotically constant behaviour in that regime, independent of $g_B$, and this asymptotic form cannot be arbitrarily contrieved (c.f. App. \ref{Ap:2VTfit}). Therefore, it is important that modified gravity theories present clear and specific predictions for the lower acceleration regime in order to be possible to differentiate them for DM-based models, such as the 2VT.


\section{Conclusion \label{Sec:disc}}

In this paper, we have revisited the two-component virial theorem (2VT) with the aim of identifying discerning features and predictions for the behaviour of gravitational systems in relation to the recent ``Acceleration Discrepancy Relation''(ADR) findings. Our main conclusions are:

{\it -- The 2VT follows approximately the ADR}, considering LTG and dSphs data.  The inferred coupling between dark and baryonic components from this relation seems to arise from their mutual equilibrium conditions. The detailed behavior of this coupling depends on the structural distribution of these components. Our work did not offer predictions on the detailed forms of such final equilibrium states, which may include a variation of the parameters $(g_D, R)$, and also dissipative mechanisms leading to different structural configurations. Indeed, the coupling between DM and baryons is a complex issue, and our previous study of this matter in the context of the 2VT indicates complementary contributions of dissipation and DM to the origin of scaling relations in astrophysical systems \citep{Rib10}. However, the overall behavior of the 2VT curve is remarkably consistent with the ADR.

{\it -- The 2VT predicts some of the main features of the ADR, such as a bending and a nearly constant behavior in the lower acceleration regime}.  The parameters $(g_D, R)$ cannot disrupt these features, only allowing for an adjustment of the ``height'' of the asymptotic behavior in the lower acceleration region, or alternatively the point at which the 2VT departures from the 1VT. 

{\it -- The somewhat ``mysterious'' dwarf spheroidal ``flattening'' behavior, if confirmed, would indicate that the 2VT provides a consistent physical description of this phenomenon} via the  dynamical equilibrium of such systems, with a highly dominated by DM component.

Finally, we point out that the ``rigidity'' of the 2VT curve is an important discriminating factor, implying that {\it this model cannot be arbitrarily contrieved to fit the data}. This level of predictive power is an important feature of the 2VT, which could, for instance, serve as a means to distinguish DM theories from non-DM (e.g., emergent gravity) theories of equilibrium systems.

\section*{Acknowledgements}
We thank the referee for useful suggestions. CCD thanks Sabine Hossenfelder for her comments on an earlier draft. ALBR thanks the support of CNPq, grant 309255/2013-9.


\appendix

\section {Rewriting the 2VT}  \label{Ap:Re2VT}

The (usual) virial theorem (hereon, 1VT) states that, for the {\it whole} system, the mean square velocity of the baryonic component is given by \citet{Bin87}:
\begin{equation}
\langle v^2 \rangle _B =  {GM_{\rm v} \over r_{B}}, \label{Eq:1VT}
\end{equation}
\noindent where $M_{\rm v}$ is the estimated virial mass of the system, and $r_B$ is the gravitational radius of the baryonic component.. It is clear that, if the virial mass is larger than the observed mass, $M_{\rm obs}$ (inferred from the observed stellar and gaseous surface mass distributions), then there is a ``hidden mass'' (or DM mass) given by the difference:

\begin{equation}
M_{\rm (D, 1VT)} = M_{\rm v} - M_{\rm obs}. \label{Eq:hidden}
\end{equation}

The 2VT gives a correction to the 1VT, and was first derived by \citet{Lim59} (c.f. App. \ref{Ap:Limber}). A simplified form of the 2VT has previously been shown to reproduce in a broad sense the scaling relations of systems at various scales \citep{Dan00}:

\begin{equation}
\langle v^2 \rangle _B = {GM_B \over r_{B}} + {4 \pi \over 3} G {\rho}_D\langle r^2 \rangle _B, \label{Eq:2VT}
\end{equation}
\noindent where the baryonic average square radius is defined by:
\begin{equation}
\langle r^2 \rangle_B \equiv {\int r^2 \rho_B(r) dV \over \int \rho_B(r) dV}. \label{Eq:rB2}
\end{equation}
\noindent In the equations above, $\rho_B(r)$ is the mass density of the baryonic component whereas $\rho_D$  is the mean density of the dark matter halo within the region containing the baryonic component, $M_B$ is the total baryonic mass within the gravitational radius.

Note that the 1VT does not address any systematic couplings between ``hidden'' and  baryonic masses, it just implies that a ``remainder'' mass must be added, by contingency,  to the dynamical equilibrium budget of the system. But in the 2VT formulation, the mean square velocity of the baryonic component {\it must be corrected in a way that it depends systematically on the dark component in which the the former is embedded}. This description of the coupling of the baryonic and DM halo is a fundamental advantage of the 2VT formulation. 

For spherically symmetric systems,  the gravitational acceleration scales for the baryonic matter and  dark matter are, respectively,
\begin{equation}
g_{\rm x} = {G M_{\rm x}(<r_B)\over r_B^2 }, \label{Eq:gB}
\end{equation}
\noindent  with $\rm x$ either referring to $B$ or $D$; $r_{B}$ is the gravitational radius of the baryonic component. In the case of axisymmetric systems, like LTGs, flattened gravitational potentials associated with finite mass systems present a Keplerian circular speed at large galactocentric distances. In this limit, we assume that Eq. \ref{Eq:gB} is approximately satisfied.

In terms of acceleration scales and mean densities, within $r_B$,
\begin{equation}
\rho_{\rm x} = {3g_{\rm x} \over 4 \pi G r_B}.  \label{Eq:rhoave}
\end{equation}

The following assumptions and definitions were used (further simplifications are described in App. \ref{Ap:Limber}):

\begin{enumerate}
\item{We define a structure parameter\footnote{This definition is slightly different than the one adopted in \citet{Dan00}.}, $R$, depending only on the properties of
the baryonic matter distribution, such that $\langle r^2 \rangle = R r^2_B$.}
\item{For rotationally supported galaxies, we make the correspondence $\langle v_{circ}^2 \rangle _B  \rightarrow  \langle v^2 \rangle_B $, where $v_{circ}$ is the circular velocity at radius $r_B$.}
\end{enumerate}

Given item (i) above, we rewrite Eq. \ref{Eq:2VT} as:
\begin{equation}
\langle v^2 \rangle _B = {GM_B \over r_{B}} + {4 \pi \over 3} G {\rho}_D Rr_B^2.\label{Eq:2VTb}
\end{equation}
\noindent On the other hand, the observed baryonic centripetal acceleration is: 
\begin{equation}
{g}_{\rm obs} = \langle v^2 \rangle _B / r_B. \label{Eq:gobs}
\end{equation}
\noindent Hence, expressing Eq. \ref{Eq:2VTb} in terms of accelerations (using Eqs.  \ref{Eq:gB}, \ref{Eq:rhoave} and \ref{Eq:gobs}), we write the 2VT prediction as:

\begin{equation}
g_{\rm obs,2VT} = g_B + Rg_D . \label{Eq:2VT-full}
\end{equation}

\section {Relation to Limber's 2VT equation} \label{Ap:Limber}

The original modification of the virial theorem proposed by Limber (Eq. 21 in  \citealt{Lim59}) for a model of a cluster of galaxies (here labeled by ``B'') with an extended, non-dissipative gaseous background (here labeled by ``D''), is given by:
\begin{equation}
\langle v^2 \rangle _B  = {GM_B \over r_{B}} 
\left [
1 +
\left (
{C_{BD}+D_{BD}  }
\right )
{M_D \over M_B}
\right ] ,\label{Eq:2VT_Limber}
\end{equation}
\noindent Our formulation of the 2VT (Eq. \ref{Eq:2VTb}) is a particular case of Limber's 2VT above. We assumed that inside the baryonic region defined by $r_B$ the DM halo was sufficiently extended, so that the average DM density inside that region did not depend on the galactocentric distance (Eq. 2 of  \citealt{Dan00}). A similar approximation can be done in Eq. (25) of Limber's paper \citet{Lim59} for the coefficient $C_{BD}$. A second assumption is that the spatial distribution of both components is similar, which implies that Limber's coefficient $D_{BD}$  is approximately zero (Eq. 26 of \citealt{Lim59}). With these approximations, our parameter $R$ depends only on the properties of
the baryonic matter distribution (c.f. App. \ref{Ap:Re2VT}), whereas in a more general formulation it would have to be re-written in terms of $C_{BD}+D_{BD}$.

\section {Characteristics of the 2VT curve in the log--log plane} \label{Ap:2VTfit}

The 2VT (Eq. \ref{Eq:2VTsum}), parametrized by a fixed value of $(g_D,R)$, represents a family of baryonic--DM systems in an arbitrary baryonic mass range. The correction for the observed acceleration, predicted by the 2VT, is given by the combined product $Rg_D$, and it is not possible to obtain separated estimates for $R$ and $g_D$. Here we illustrate how the 2VT curve is affected by these quantities, by fixing one and varying the other, and vice-versa. 

\begin{figure} 
\centering
\includegraphics[width=0.49\columnwidth]{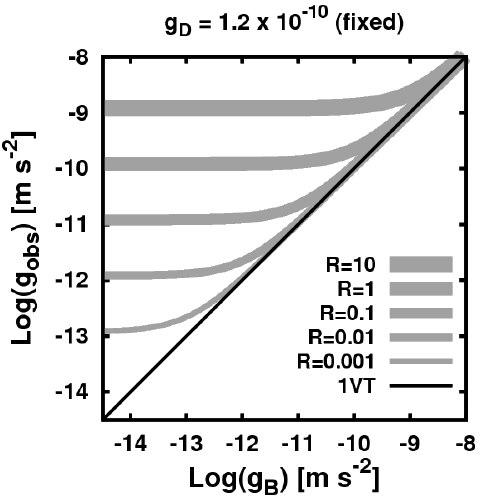}
\includegraphics[width=0.49\columnwidth]{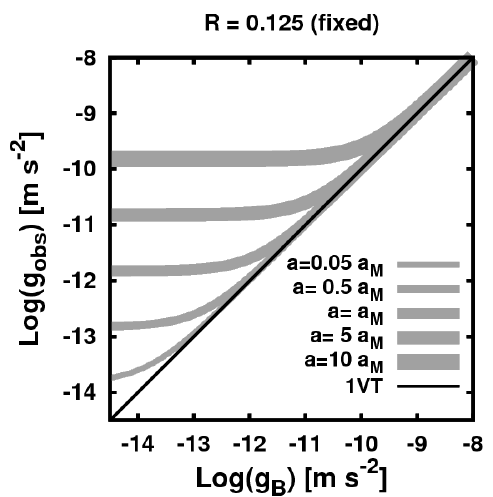}
\caption{The acceleration associated with the baryonic matter ($g_{\rm B}$) vs. the observed acceleration predicted by the 2V (Eq. \ref{Eq:2VTsum}). Left panel shows 2VT curves with $g_D$ fixed and varying $R$ values, whereas the right panel shows the other way around  (their corresponding values are indicated in the legends). These panels illustrate how these parameters affect the departure from the 1VT  (the line of identity) in a similar fashion.\label{Fig:2VT-App}}
\end{figure}

In Fig. \ref{Fig:2VT-App}, left panel, we fix $g_D$ to Milgrom's acceleration scale (Eq. \ref{milgrom_am}) and vary the parameter $R$, whereas in the right panel, we fix $R=0.125$ and vary $g_D$. As expected, both parameters produce similar effects on the 2VT. Clearly, the height of the asymptotic part of the 2VT in the lower acceleration region is solely regulated by the departure from the 1VT.  It is important to notice that the parameters $(g_D,R)$ cannot disrupt this rigid form of the 2VT, only allowing for an adjustment of the height of the asymptotic  behaviour of the curve or, alternatively, the point at which the 2VT departures from the 1VT.

\bibliographystyle{mnras}
\bibliography{CCDantas_2017}

\bsp	
\label{lastpage}
\end{document}